\documentclass{jps-cp}
\usepackage{txfonts} %Please comment out this line unless the txfonts package is availabe in your LaTeX system.

\usepackage{graphicx,color}
\usepackage{amsmath,amssymb,bm,ascmac,calc}

\newcommand{\bra}{\langle}  % added by HN
\newcommand{\ket}{\rangle}  % added by HN
\newcommand{\vect}[1]{\boldsymbol{#1}}  % added by HN

\title{Anatomy of charge radii of Ca nuclei in self-consistent approaches
      with Skyrme and Fayans functionals}

\author{H. \textsc{Nakada}$^{1,2}$, T. \textsc{Inakura}$^{3}$ and N. \textsc{Hinohara}$^{4,5}$}

\inst{
  $^{1}$Department of Physics, Graduate School of Science, Chiba University,
  Inage, Chiba 263-8522, Japan\\
  $^{2}$Research Center for Nuclear Physics, The University of Osaka,
  Ibaraki, Osaka 567-0047, Japan\\
  $^{3}$Office of Institute Strategy, Institute of Science Tokyo,
  Meguro, Tokyo 152-8550, Japan\\
  $^{4}$Center for Computational Sciences, University of Tsukuba,
  Tsukuba, Ibaraki 305-8577, Japan\\
  $^{5}$Faculty of Pure and Applied Sciences, University of Tsukuba,
  Tsukuba, Ibaraki 305-8571, Japan
}

\email{nakada@faculty.chiba-u.jp}

\recdate{September 30, 2026}

\abst{
  It has been known that the $^{40-48}$Ca nuclei
  have anomalous $N$-dependence of their charge radii.
  While difficult to reproduce in most self-consistent calculations,
  the parabolic behavior in $^{42,44,46}$Ca is reproduced
  with the Fayans functional,
  although it tends to overestimate the charge radii
  at the neutron-deficient $^{38}$Ca.
  We investigate the origin of the $N$-dependence with the Fayans functional
  by decomposing it into radial contributions,
  and trace it back to the single-particle potentials
  and neutron pair densities.
  The results are compared with those of a Skyrme functional SkM$^\ast$,
  which is considered a representative of the usual functionals.
  It is found that, in the results of the Fayans functional,
  the origin is shared between the parabolic behavior in $^{42,44,46}$Ca
  and the erroneous  enhancement of the charge radii at $^{38}$Ca.
  To fully resolve the puzzle with respect to the parabolic charge radii
  in $^{42,44,46}$Ca,
  it is vital to reproduce the neutron-deficient region simultaneously.
}

\kword{Charge radii, Charge distribution, Calcium isotopes,
  Density functional theory, Fayans functional, Skyrme functional}

\begin{document}
\maketitle

\section{Introduction}

%The \verb|minipage| environment can be used to place figures horizontally.

While nuclear radii are proportional to $A^{1/3}$ at the zeroth approximation
that manifests the saturation~\cite{Hey04},
deviation from this $A^{1/3}$-rule is a source of rich physics:
shape evolution~\cite{BM1},
nuclear halos near the drip line~\cite{Tan13},
and irregularities such as kinks (and anti-kinks)~\cite{Nak19}.
The density-functional theory (DFT) offers variational calculations
of the nuclear properties via the local density $\rho(\vect{r})$~\cite{Colo20},
in which the nuclear mean-square (ms) radius is a basic quantity
carrying the simplest information of the density,
\begin{equation} \bra r^2\ket_\tau
  = \frac{1}{N_\tau} \int r^2\,\rho_\tau(\vect{r})\,d^3r\,,
  \quad(\tau=\mathrm{p},\mathrm{n})
  \label{r2&rho}\end{equation}
where $N_\tau=Z$ ($N$) for $\tau=\mathrm{p}$ ($\mathrm{n}$).
In particular, the Hartree-Fock-Bogoliubov (HFB) theory
is a suitable framework for studying the nuclear radii,
which respects the self-consistency
between the single-particle (s.p.) potential and density
in the presence of the pair correlations.

Among the nuclear radii, the charge radii supply precise data~\cite{AM13},
measured by the electromagnetic probes.
Their absolute values have been obtained
from electron elastic scattering~\cite{DeVries},
which has long been restricted to stable nuclei
but has been extended to unstable nuclei via the SCRIT technology~\cite{SCRIT}.
The differential ms charge radii have been accurately extracted
from the isotope shifts,
for which abundant data are provided on unstable nuclei
by the laser spectroscopy experiments~\cite{LaserSpectroscopy}.
The isotopic difference of charge radii in magic-$Z$ nuclei
has attracted special interest.
Because occupied proton orbitals are common in the magic-$Z$ nuclei,
it likely reflects variation of the s.p. wave functions
due to the attraction from neutrons.
The origin and the size of the isotopic dependence are subject
to the effective nucleonic interaction.
For instance, a role of the spin-orbit ($\ell s$) potential
in the kink at $^{208}$Pb has been argued~\cite{Sharma,RF95,NI15}.

\begin{figure}[tbh]
  \begin{center}
  \includegraphics[scale=0.4]{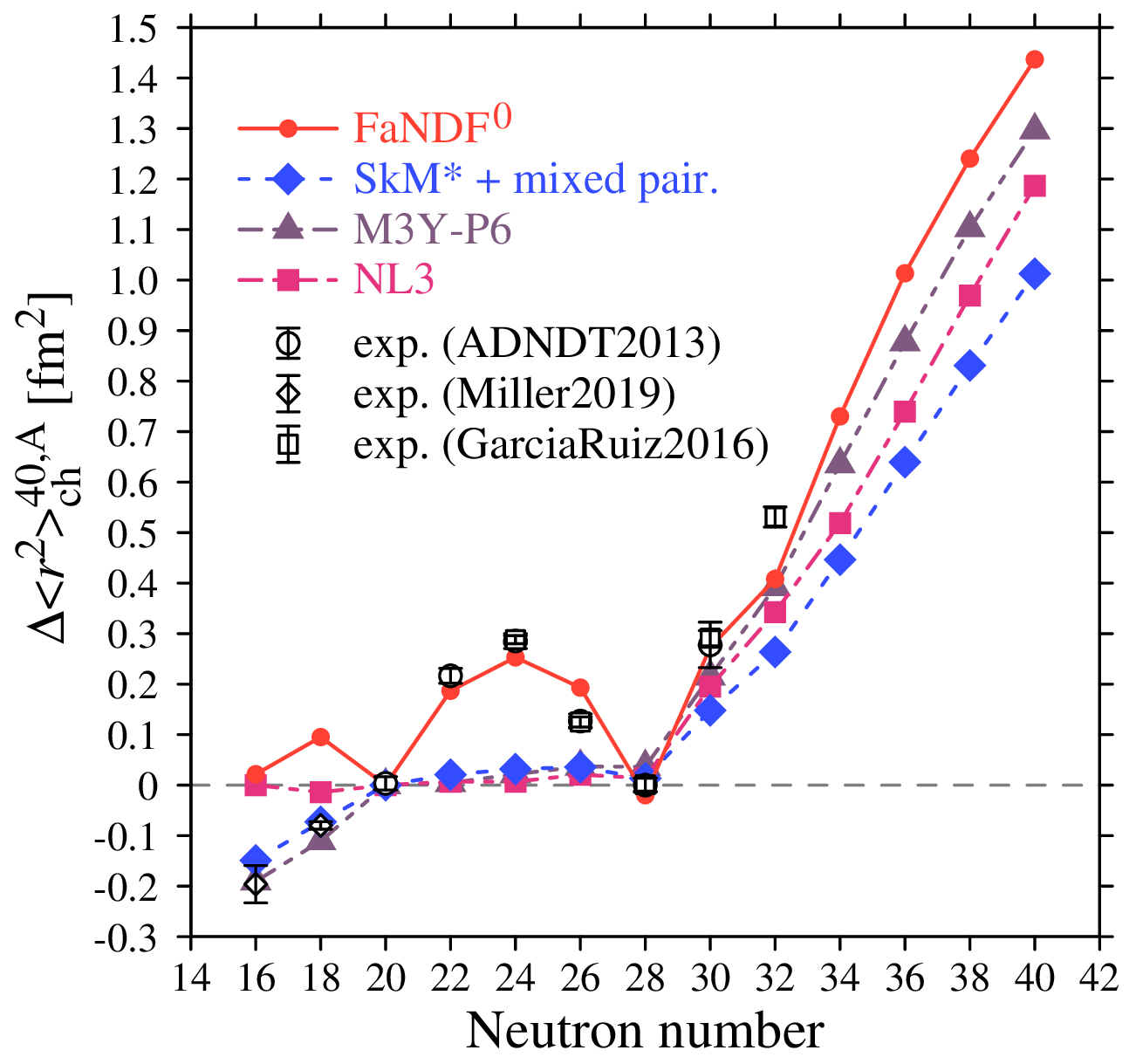}
  \caption{
    Calculated and measured differential charge radii in Ca nuclei.
    Experimental data are taken from Refs.~\cite{AM13,GBB16,MMK19},
    and the NL3 results from Ref.~\cite{LRR99}.
    Quoted from Ref.~\cite{IHN24} with modification under permission
    (Copyright by the American Physical Society).}
  \label{drch_Ca}
  \end{center}
\end{figure}

Charge radii in stable Ca nuclei pose a long-standing problem.
While the measured charge radii of even-$N$ Ca
behave parabolically in $20\leq N\leq 28$,
this behavior is difficult to reproduce by the HFB calculations,
in which pairing is active among neutrons except at $A=40$ and $48$,
but not among protons because of the $Z=20$ magicity.
Perturbative effects hardly account for this behavior, either~\cite{Co22}.
There have been suggestions that the radii correlate
with the extraordinarily strong $E2$ transitions~\cite{BM21}.
This correlation may be linked to excitation of protons
out of the $sd$-shell~\cite{Cau01}.
On the other hand, as illustrated in Fig.~\ref{drch_Ca}
in terms of $\mathit{\Delta}\bra r^2\ket_\mathrm{ch}^{A_0,A}
:= \bra r^2\ket_\mathrm{ch}(\mbox{$^A$Ca})
- \bra r^2\ket_\mathrm{ch}(\mbox{$^{A_0}$Ca})$,
where $\bra r^2\ket_\mathrm{ch}$ denotes the ms charge radius,
the Fayans density functional~\cite{Fay01}
yields a parabolic behavior as an effect of its specific pairing,
without breaking the $Z=20$ core.
It is noted that the Fayans functional gives a kink at $^{208}$Pb,
and at $^{132}$Sn prior to the experiment~\cite{GRB19}
together with Ref.~\cite{Nak15}.
However, to our best knowledge,
the Fayans functional that reproduces the parabolic behavior in $^{40-48}$Ca
fails to describe the decrease of the charge radii from $^{40}$Ca to $^{38}$Ca.
We investigate how the Fayans functional makes difference from usual ones
in the charge radii in the Ca nuclei~\cite{IHN24}
(see also \cite{IHN24_correct}),
aiming to help address the correct physics via future studies.
For this purpose,
we use the FaNDF$^0$ parameter-set of the Fayans functional,
and the SkM$^\ast$ functional combined with the mixed pairing channel.
Other available parameter-sets~\cite{BT22,Rein24,NCRS26}
of the Fayans functional do not yield significant difference,
and SkM$^\ast$ is regarded as a representative of the normal functionals.
Numerical calculations have been implemented
with the code \texttt{HFBRAD}~\cite{BD05}.
The charge radii and densities have been calculated
with the corrections, following Ref.~\cite{KS21}.
We focus on the even-$N$ Ca isotopes
to avoid influence of weak deformation due to a quasiparticle.
Besides its degree,
the rapid increase of the charge radii in $N>28$
is qualitatively described with all functionals in Fig.~\ref{drch_Ca}.

\section{Isotopic difference of density distribution}

As long as $\rho_\tau(r)$ is taken to be a principal variable,
not just the ms radii but also $\rho_\tau(r)$ itself
should be correctly described.
We compare results of FaNDF$^0$ and SkM$^\ast$
for the charge densities, point-proton densities,
and their isotopic difference defined as
\begin{equation}
  \bar{\mathit\Delta}\rho_{{\rm ch}/{\rm p}}^{A_0,A}(r)
  := (4\pi r^2) r^2 \left[\rho_{{\rm ch}/{\rm p}}(r;\mbox{$^{A}$Ca})
        - \rho_{{\rm ch}/{\rm p}}(r;\mbox{$^{A_0}$Ca})\right]\,,
\end{equation}
whose integration over $r$ is proportional to the differential ms radius
$\mathit{\Delta}\bra r^2\ket_{{\rm ch}/{\rm p}}^{A_0,A}$.
Taking $^{40}$Ca as the reference nuclide (\textit{i.e.}, $A_0=40$),
$\rho_{\rm ch}(r)$ and $\bar{\mathit\Delta}\rho_{\rm ch}^{A_0,A}(r)$
are compared with those extracted
from the $(e,e)$ experiments through the Fourier-Bessel (FB) analysis
or assuming the three-parameter Fermi (3pF) form
of $\rho_{\rm ch}(r)$~\cite{DeVries}.

\begin{figure}[tbh]
  \begin{center}
  \includegraphics[scale=0.33]{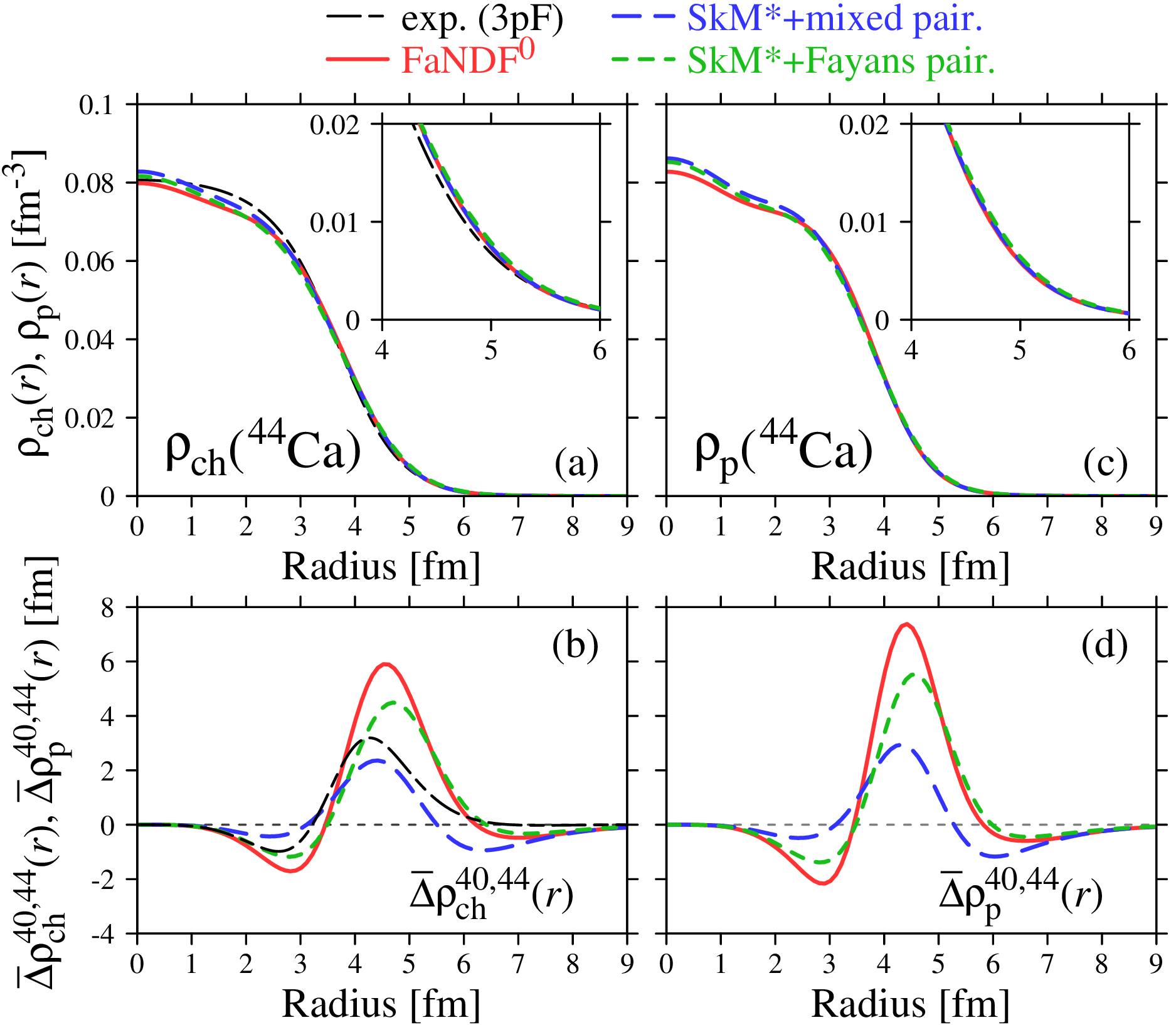}
  \caption{
    (a) $\rho_{\rm ch}(r;\mbox{$^{44}$Ca})$,
    (b) $\bar{\mathit\Delta}\rho_{\rm ch}^{40,44}(r)$,
    (c) $\rho_{\rm p}(r;\mbox{$^{44}$Ca})$ and
    (d) $\bar{\mathit\Delta}\rho_{\rm p}^{40,44}(r)$
    calculated with FaNDF$^0$, SkM$^\ast$\,+\,mixed pairing,
    and SkM$^\ast$\,+\,Fayans pairing.
    In (a,\,b), experimentally extracted ones with 3pF~\cite{DeVries}
    are shown for comparison.
    Quoted from Ref.~\cite{IHN24}.}\label{fig:drho_Ca44}
  \end{center}
\end{figure}

For $^{48}$Ca, at which the pairing vanishes,
the calculated $\rho_{{\rm ch}/{\rm p}}(r)$
and $\bar{\mathit\Delta}\rho_{{\rm ch}/{\rm p}}^{40,48}(r)$
resemble between SkM$^\ast$ and FaNDF$^0$, and are almost consistent
with the FB and 3pF experimental densities~\cite{IHN24}.
Figure~\ref{fig:drho_Ca44} depicts
$\rho_{{\rm ch}/{\rm p}}(r;\mbox{$^{44}$Ca})$
and $\bar{\mathit\Delta}\rho_{{\rm ch}/{\rm p}}^{40,44}(r)$.
The result of SkM$^\ast$ but substituting the pairing channel
with that of FaNDF$^0$ is also presented.
We find that $\bar{\mathit\Delta}\rho_{{\rm ch}/{\rm p}}^{40,44}(r)$
of SkM$^\ast$
is similar to $\bar{\mathit\Delta}\rho_{{\rm ch}/{\rm p}}^{40,48}(r)$,
and in good agreement with the experiments at $r\lesssim 5\,\mathrm{fm}$,
though the dip in the outer region ($r\sim 6\,\mathrm{fm}$)
offsets the contribution of the peak at $r\sim 4\,\mathrm{fm}$.
In contrast, $\bar{\mathit\Delta}\rho_{\rm ch}^{40,44}(r)$ of FaNDF$^0$
is not in agreement with the experimental one,
with a pronounced peak (and dip) at a shifted position;
agreement in $\bra r^2\ket_{\rm ch}$ with FaNDF$^0$ after integration
seems accidental, not originating from agreement in $\rho_{\rm ch}(r)$.
However, we should not be too conclusive,
since there could be room to question
the accuracy of the experimental $\rho_{\rm ch}(r)$.
No qualitative difference is found
between $\rho_\mathrm{p}(r)$ and $\rho_\mathrm{ch}(r)$,
besides the peak height and dip depth.
Hereafter we discuss $\rho_\mathrm{p}(r)$.

\begin{figure}[tbh]
%  \includegraphics[scale=0.35]{figs/proton.density.52Ca.60Ca.FaNDF0.SkMs.paper.eps}%\vspace*{-9cm}
%  \centerline{\includegraphics[scale=0.35]{figs/proton.density.52Ca.60Ca.FaNDF0.SkMs.ARISproceedings.eps}}
%  \centerline{\includegraphics[scale=0.33]{figs/proton.density.52Ca.60Ca.FaNDF0.SkMs.ARISproceedings.eps}}
  \centerline{\includegraphics[scale=0.33]{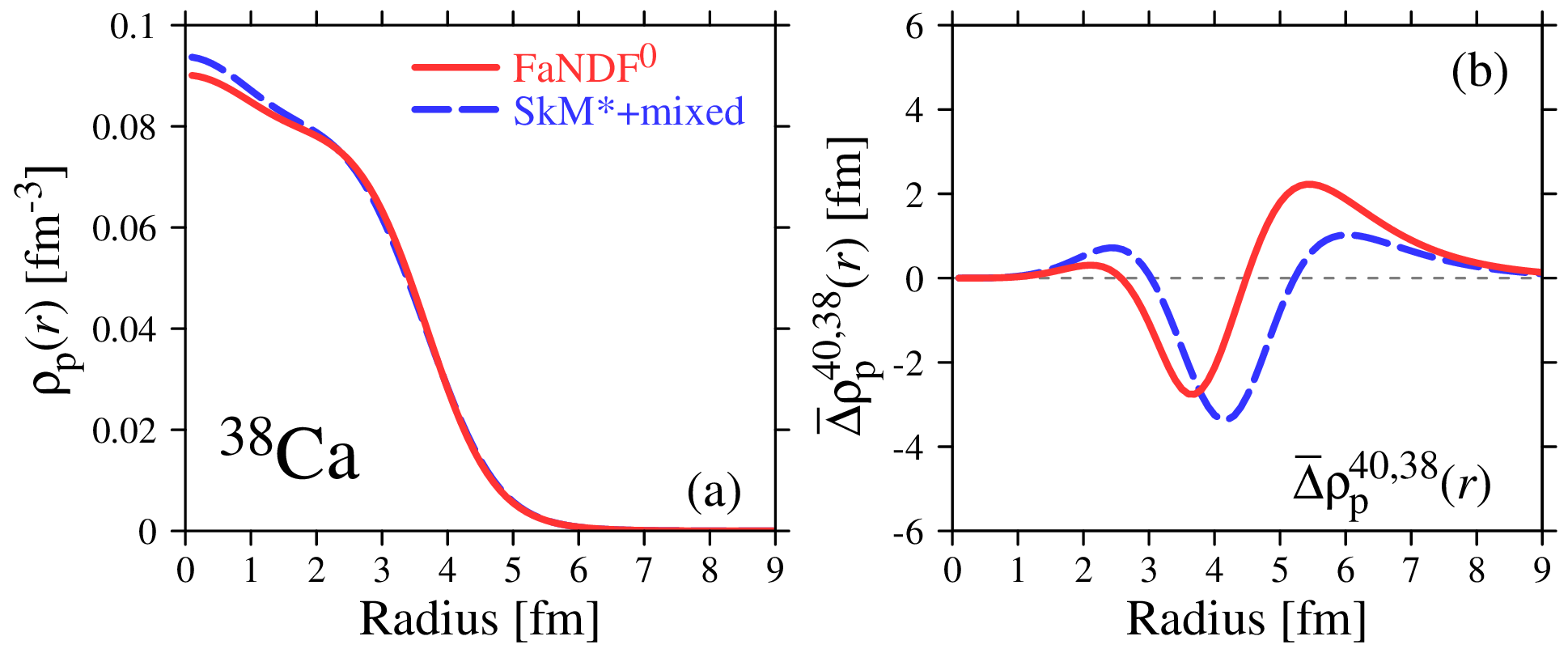}}
  \caption{
    (a) $\rho_{\rm p}(r;\mbox{$^{38}$Ca})$ and
    (b) $\bar{\mathit\Delta}\rho_{\rm p}^{40,38}(r)$
    calculated with FaNDF$^0$ and SkM$^\ast$\,+\,mixed pairing.
    }\label{fig:drho_Ca38}
\end{figure}

In Fig.~\ref{fig:drho_Ca38},
$\rho_{\rm p}(r)$ and $\bar{\mathit\Delta}\rho_{\rm p}(r)$ are displayed
for $^{38}$Ca.
It should be noted that difference between FaNDF$^0$ and SkM$^\ast$
well corresponds to that at $^{44}$Ca.

\section{Difference in single-particle potential}

\begin{figure}[tbh]
  \vspace*{-0.5cm}
  \begin{center}
  \includegraphics[scale=0.35]{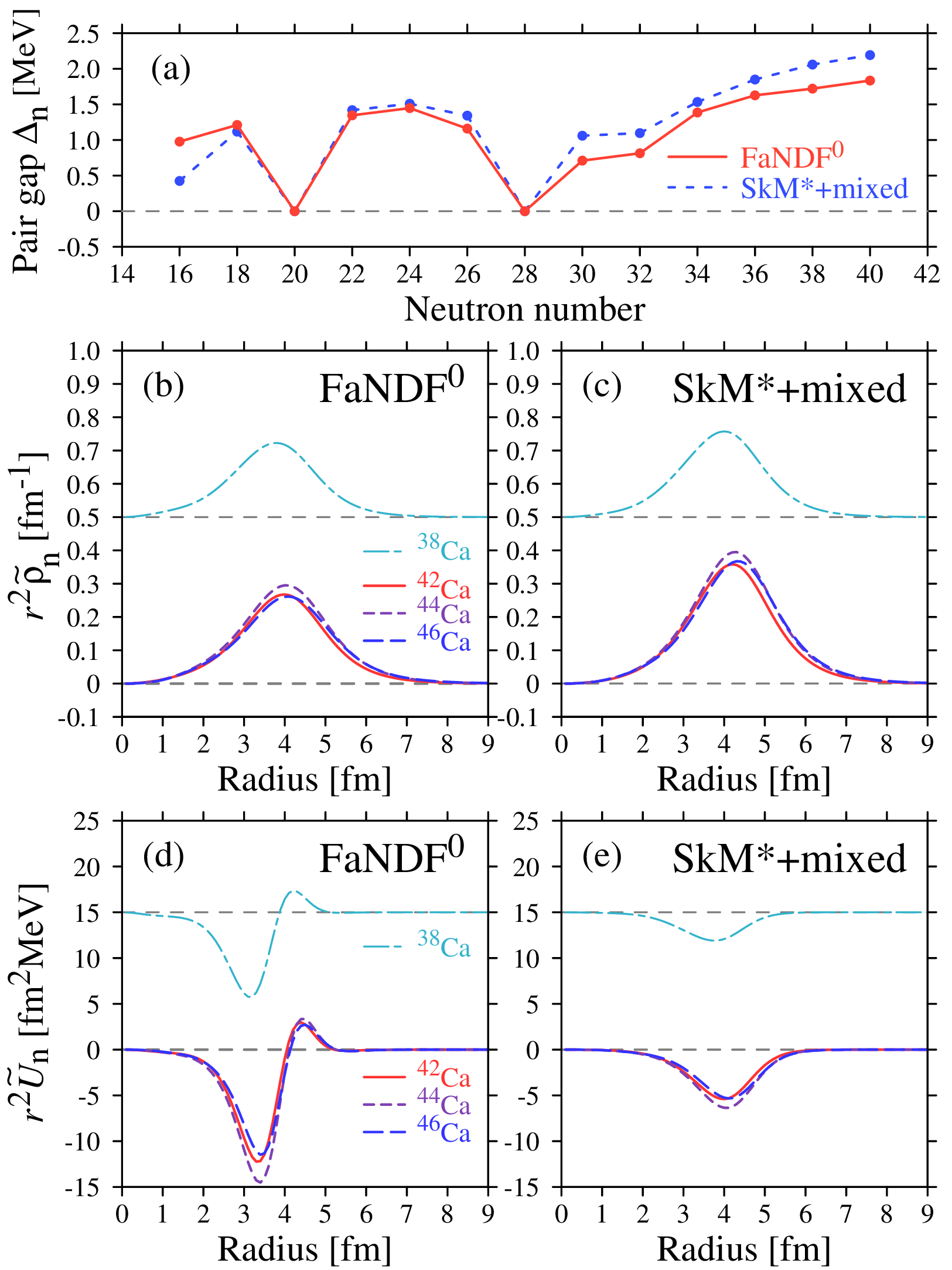}
  \caption{
    (a) Pairing gaps $\Delta_{\rm n}$,
    (b,\,c) pair densities $r^2\tilde\rho_{\rm n}(r)$ and
    (d,\,e) pair potentials $r^2\tilde U_{\rm n}(r)$ of neutrons
    calculated with FaNDF$^0$ and SkM$^\ast$\,+\,mixed pairing.
    In (b\,--\,e), the lines are properly shifted in the vertical direction.
    }\label{fig:n-pair}
  \end{center}
\end{figure}

\begin{figure}[tbh]
  \vspace*{-0.5cm}
  \begin{center}
  \includegraphics[scale=0.3]{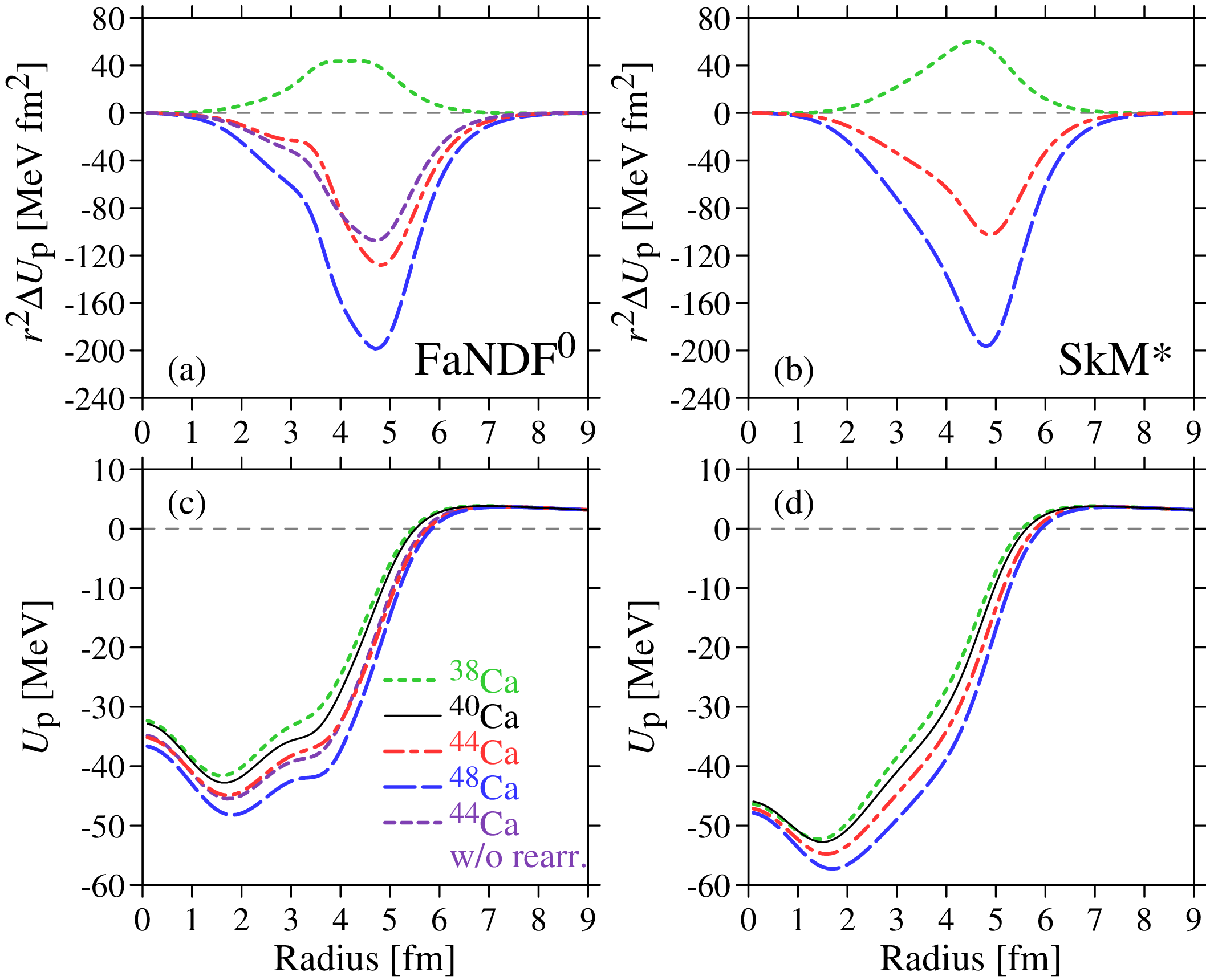}
  \caption{
    (a,\,b) Difference of central potential $r^2\mathit\Delta U_{\rm p}(r)$
    relative to $^{40}$Ca,
    (c,\,d) central potential $U_{\rm p}(r)$,
    of protons,
    calculated with FaNDF$^0$ and SkM$^\ast$.
    }\label{fig:dUp}
  \end{center}
\end{figure}

We have traced back the origin of the Fayans-functional's
peculiar $\bra r^2\ket_{{\rm ch}/{\rm p}}$
to $\rho_{{\rm ch}/{\rm p}}(r)$.
Because $\rho_{\rm p}(r)$ is determined by the proton s.p. potential $U_{\rm p}$,
its isotopic difference should be traced back
to the isotopic difference of $U_{\rm p}$.
As the peculiar $\bra r^2\ket_{\rm p}$ of the Fayans functional
has been confirmed to be an effect of the rearrangement potential
due to its specific pairing channel~\cite{NCRS26,RN17},
it should be further traced back to effects of the neutron pairing.
In Fig.~\ref{fig:n-pair}, the pair densities ($\tilde\rho_{\rm n}$),
pair potentials ($\tilde U_{\rm n}$),
as well as the density-averaged neutron pairing gaps,
$\Delta_{\rm n} := -\frac{1}{N} \int d^3r\,\tilde U_{\rm n}(r)\,\rho_{\rm n}(r)$,
are comparatively displayed for SkM$^\ast$ and FaNDF$^0$.
The proton potential affected by the neutron pairing
is then compared in Fig.~\ref{fig:dUp}.

Despite similar values of $\Delta_{\rm n}$,
FaNDF$^0$ provides peculiar $r^2 \tilde{U}_\mathrm{n}(r)$,
having a deep valley near surface ($r\sim 3.5\,\mathrm{fm}$),
a bump beyond surface ($r\sim 4.5\,\mathrm{fm}$),
and suddenly vanishing after the bump.
%Through the pair density $\tilde\rho_\mathrm{n}(r)$
%and the induced rearrangement contribution,
Affecting the pair density $\tilde\rho_\mathrm{n}(r)$
and inducing rearrangement contribution,
this $\tilde{U}_\mathrm{n}(r)$ produces
a peculiar proton potential $U_\mathrm{p}(r)$,
as more visible in its isotopic difference $r^2 \mathit\Delta U_\mathrm{p}(r)$
typified by the slope change near the surface ($r\sim 3.5\,\mathrm{fm}$).
Thus, $\bra r^2\ket_{{\rm ch}/{\rm p}}$ in $^{42,44,46}$Ca
specific to the Fayans functionals
has been traced back and decomposed
into the $r$-dependence of the neutron pairing.
It is remarked that this mechanism from neutron pairing
to the proton or charge radius
is common between the stable $^{42,44,46}$Ca and the neutron-deficient $^{38}$Ca.
Therefore, when the Fayans functional yielding the parabolic ms charge radii
in the stable Ca is adopted,
the $^{38}$Ca radius quite likely increases from $^{40}$Ca,
in contradiction to the data.
It is also pointed out in the recent analysis~\cite{NCRS26}
that, with the Fayans functional form,
the neutron pairing yields repulsive rearrangement contribution
to the proton potential,
which is also shared between $^{42,44,46}$Ca and $^{38}$Ca.
To fully resolve the puzzle of the Ca radii,
it is significant to reproduce the radii in the stable region
and the neutron-deficient region simultaneously.

\section{Summary}

The Fayans functional has been known to reproduce the parabolic $N$-dependence
of the charge radii in $^{40-48}$Ca,
although it hardly reproduces the charge radius of $^{38}$Ca.
To comprehend its mechanism and assess the physics rationality of the results,
we decompose the ms radii in terms of the density distributions,
and compare those given by FaNDF$^0$
with those by the Skyrme functional SkM$^\ast$ combined with a normal pairing.
In the isotopic difference $\bar{\mathit\Delta}\rho_{\rm ch}^{40,44}(r)$,
the FaNDF$^0$ result is not consistent with available experimental data,
being in agreement only with the integrated value,
although the accuracy of the $\bar{\mathit\Delta}\rho_{\rm ch}^{40,44}(r)$ data
remains questionable.
We further trace back $\rho_{{\rm ch}/{\rm p}}(r)$ to the proton potential,
and to the neutron pair potential and density,
disclosing how the specific pairing channel of the Fayans functional
gives rise to the peculiar $N$-dependence of the charge radii.
It is noticed that the mechanism is common
between the stable Ca and the neutron-deficient $^{38}$Ca,
indicating that the success in $^{42,44,46}$Ca
and overestimation at $^{38}$Ca have the same origin.
To fully resolve the puzzle in the charge radii of Ca isotopes,
simultaneous description of stable and neutron-deficient Ca
is highly desired.

\section*{Acknowledgment}

The authors are grateful to T.~Naito for useful comments and discussions.
%This work was supported by the JSPS KAKENHI
%(Grant Nos. JP19KK0343, JP20K03964 and JP24K07012)
This work was supported by the JSPS KAKENHI (Grant No. JP24K07012)
and the JST ERATO (Grant No. JPMJER2304).


\begin{thebibliography}{9}
\bibitem{IHN24_correct}
  There were errors on the rhs's of Eqs.~(18), (19), (30), and (B8)
  of Ref.~\cite{IHN24}.
  The overall coefficients should be
  $\epsilon_F/3\rho_\mathrm{sat}$ in Eqs.~(18), (30), and (B8)
  and $4\epsilon_F/3\rho_\mathrm{sat}$ in Eq.~(19).
  Note that we define the pair density as in Ref.~\cite{BD05}.
\end{thebibliography}
\end{document}